\documentclass[%
 reprint,
 superscriptaddress,
 amsmath,amssymb,
 aps,
 prl,
 longbibliography,
lengthcheck,%
]{revtex4-1}

\usepackage{graphicx}% Include figure files
\usepackage{dcolumn}% Align table columns on decimal point
\usepackage{bm}% bold math
\usepackage{color}
\usepackage{ulem}
\usepackage{hyperref}

\begin{document}

%\preprint{APS/123-QED}

\title{
Theory of giant skew scattering by spin cluster
}

\author{Hiroaki Ishizuka}
\affiliation{
Department of Applied Physics, The University of Tokyo, Bunkyo, Tokyo, 113-8656, JAPAN 
}

\author{Naoto Nagaosa}
\affiliation{
Department of Applied Physics, The University of Tokyo, Bunkyo, Tokyo, 113-8656, JAPAN 
}
\affiliation{
RIKEN Center for Emergent Matter Sciences (CEMS), Wako, Saitama, 351-0198, JAPAN
}

\date{\today}

\begin{abstract}
  Skew scattering of electrons induced by a spin cluster is studied theoretically focusing on metals with localized magnetic moments. The scattering probability is calculated by a non-perturbative $T$ matrix method; this method is valid for arbitrary strength of electron-spin coupling. We show the scattering of electrons by a three-spin cluster produces a skew angle of order $0.1\pi$ rad when the electron-spin coupling is comparable to the bandwidth. This is one or two orders of magnitude larger than the usual skew angle by an impurity with spin-orbit interaction. Systematic analysis of the scattering probability of one-, two-, and three-spin clusters show that three spins are necessary for skew scattering. We also discuss the relation between anomalous/spin Hall effects and the spin chiralities; we find that the spin Hall effect requires three spins while it is related to the vector spin chirality defined by a pair of spins. The relevance of these results to the large extrinsic anomalous and spin Hall effects in noncentrosymmetric and/or frustrated magnets is also discussed.
\end{abstract}

\pacs{
}% PACS, the Physics and Astronomy
% Classification Scheme.

\maketitle

Anomalous and spin Hall effect reflects rich physics related to the quantum nature of electrons such as Berry phase and electron scattering by impurities~\cite{Nagaosa2010,Sinova2015,Maekawa2017}. Traditionally, the microscopic mechanisms of these transport phenomena are classified into two groups: intrinsic and extrinsic mechanisms. The intrinsic mechanism of anomalous Hall effect (AHE)~\cite{Karplus1954} is related to the Berry curvature of electronic bands~\cite{Xiao2010}. Later it was realized that the same mechanism also produces intrinsic spin Hall effect (SHE)~\cite{Murakami2003,Sinova2004}. More recently, it was pointed out that the scalar spin chirality of ordered magnetic moments also contributes to the AHE~\cite{Ye1999,Ohgushi2000,Shindou2001}. This mechanism is thought to be responsible for the intrinsic AHE in ordered phases of magnets with non-coplanar magnetic order, such as in pyrochlore~\cite{Taguchi2001} and kagome~\cite{Nakatsuji2015} magnets, and in chiral magnets~\cite{Neubauer2009,Kanazawa2011}. On the other hand, the extrinsic mechanisms of AHE are related to impurity scattering. Several mechanisms are known for single non-magnetic~\cite{Smit1958,Berger1970} or magnetic~\cite{Kondo1962,Levy1987,Yamada1993} impurities; they also contribute to the SHE~\cite{Dyakonov1971,Hirsch1999}. While a variety of mechanisms are known, in three-dimensional materials, the Hall angle of anomalous Hall conductivity $\sigma_{xy}^\text{(AHE)}$ is usually small compared to the longitudinal conductivity $\sigma_{xx}$. Typically $\sigma_{xy}^\text{(AHE)}/\sigma_{xx}=10^{-3}-10^{-2}$ regardless of the mechanism~\cite{Onoda2008}.

\begin{figure}
  \includegraphics[width=\linewidth]{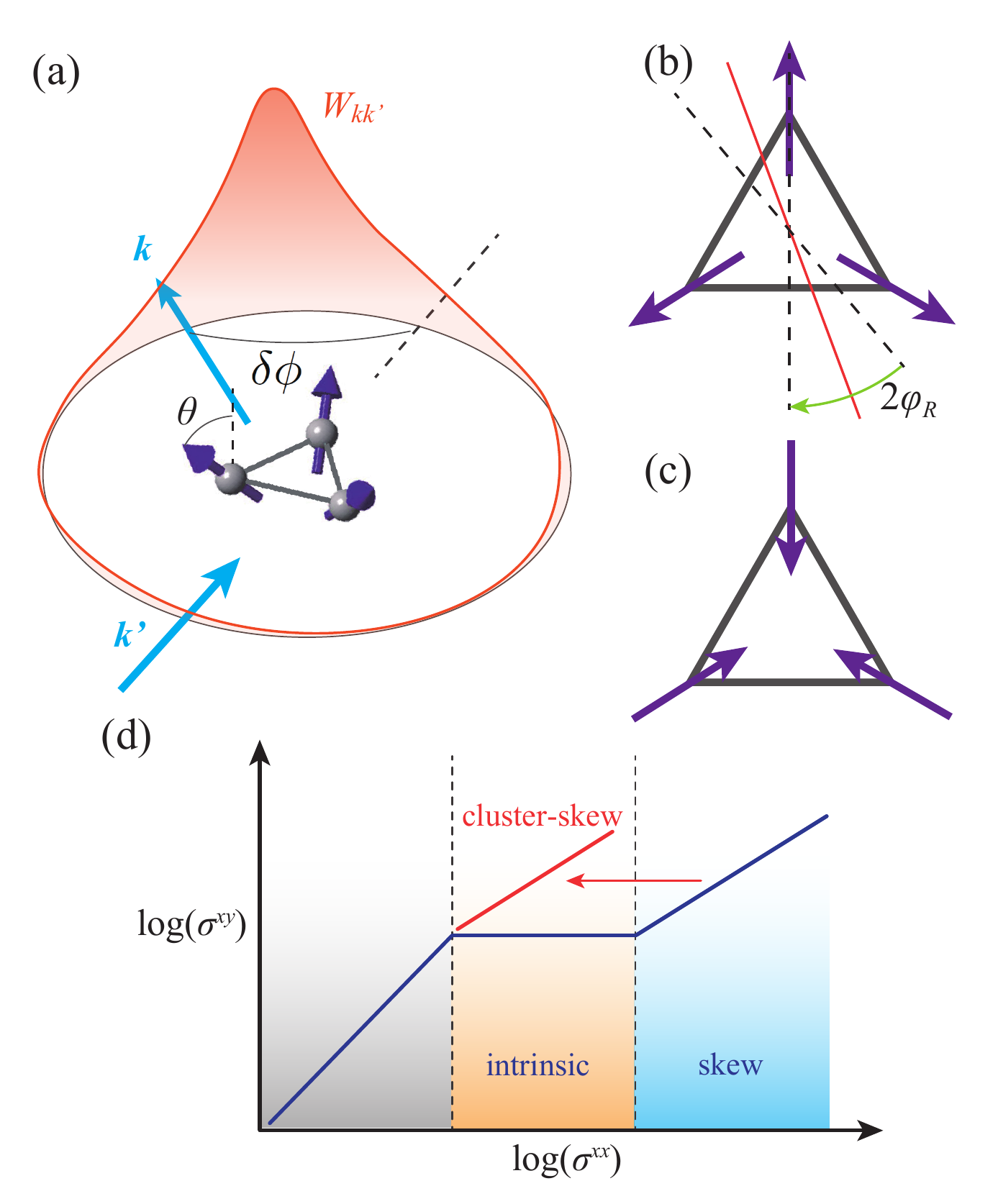}
  \caption{Schematic figure of a three-spin cluster and skew scattering. (a) Schematic figure of the electron scattering by a three-spin cluster. The blue arrows show the incoming ($\bm k'$) and outgoing ($\bm k$) electrons and the orange curve surrounding the spin cluster is the scattering rate $W_{\bm k\bm k'}$ for the outgoing electrons for $\bm k'$; we abbreviate the spin indices of the electrons. The skew scattering makes the scattering rate asymmetric with respect to the incident direction shown by the dashed line. $\theta$ in (a) is the canting angle of the three spins. (b) and (c) are respectively the top view of the three-spin cluster canted outward (b) and inward (c). See the main text for details. (d) Schematic figure of the scaling relation of anomalous Hall effect.}
  \label{fig:intro:skew}
\end{figure}

In the case of the extrinsic mechanisms, the small Hall angle is related to the necessity of the spin-orbit interaction. All extrinsic mechanisms by single impurity require spin-orbit interaction. For example, the major contribution is believed to be the skew scattering, where the electrons are scattered asymmetrically by the spin-orbit interaction of the impurity. In a typical ferromagnet, this spin-orbit interaction is thought to be a weak perturbation compared with the energy scale of the hybridization between the resonance state and the conduction electrons. Therefore, the skew angle of the scattering is typically very small, which only produces a small AHE. In contrast, such limitation does not apply to the skew scattering by multiple scatterers. The scattering by multiple magnetic scatterers also contributes to AHE; the AHE is directly related to the scalar spin chirality of impurity spins~\cite{Tatara2002}. Later, it was shown that this AHE is an extrinsic AHE by the skew scattering related to the three-spin scattering~\cite{Denisov2016,Ishizuka2017}. In addition, a mechanism related to the vector spin chirality also contributes to the AHE in certain cases~\cite{Taguchi2009,Yi2009,Zhang2018,Ishizuka2018}. These studies so far focus on the weak-coupling limit, in which the impurities are treated as perturbations; studies on related phenomena in the strong-coupling cases are limited to several numerical works~\cite{Yi2009,Ishizuka2013,Chern2014,Ishizuka2013b}. On the other hand, experimentally, the strong-coupling cases are often realized in transition-metal materials, e.g., in Mn compounds~\cite{Zener1951,Anderson1955}. However, much less is known about the multiple-spin scattering when the electron-spin coupling is strong.

In this work, we systematically study the skew scattering by multiple spins using a $T$-matrix approach. The $T$ matrix is calculated by a Green's function method for the Anderson impurity model. From the $T$ matrix, we study the skew scattering by a three-spin cluster scattering. We find that the three-spin cluster causes a skew scattering with a large skew angle in the order of $0.1\pi$ rad when the electron-spin coupling is strong. This skew angle is 10-100 times larger than the typical scattering angle of the skew scattering by single impurity. The skew scattering may produce a large Hall angle in the magnetic metals if the scalar spin chirality of fluctuating spins remains finite. In addition, we find the spin clusters also produce a large spin-dependent skew scattering. We further discuss that the skew scattering is related to the net vector spin chirality of the three pairs of spins. This spin-dependent skew scattering is expected to produce a large extrinsic spin Hall effect, which is potentially relevant to the spin Hall effect in spin glasses~\cite{Jiao2018}.

\section{Results}

\vspace{.4cm}
\noindent
{\bf Model\\}\hspace{.3cm}
We here study the $T$ matrix of a triangular lattice model with three impurity sites subject to Zeeman field. The Hamiltonian is
\begin{subequations}
\begin{align}
H=&H_f+H_c+H_{fc}+H_{cf},\label{eq:Hamil}\\
H_f=&-J\sum_{i=0,1,2} \bm S_i\cdot f_i^\dagger \bm\sigma f_i,\\
H_c=&\sum_{\bm k}\varepsilon_{\bm k} c_{\bm k}^\dagger c_{\bm k},\\
H_{fc}=& -\frac{V}{\sqrt N}\sum_{\substack{i=0,1,2\\\bm k,\sigma}}\gamma_{i\bm k}\, f_{i\sigma}^\dagger c_{\bm k \sigma},\\
H_{cf}=& -\frac{V}{\sqrt N}\sum_{\substack{i=0,1,2\\\bm k,\sigma}}\gamma^\ast_{i\bm k}\, c_{\bm k \sigma}^\dagger f_{i\sigma},
\end{align}
\end{subequations}
where $c_{\bm k\sigma}$ and $f_{\bm k\sigma}$ ($c_{\bm k\sigma}^\dagger$ and $f_{\bm k\sigma}^\dagger$) are respectively the annihilation (creation) operator of itinerant and localized electrons, $\vec \sigma\equiv(\sigma^x,\sigma^y,\sigma^z)$ is the vector of Pauli matrices $\sigma^a$ ($a=x,y,z$), $c_{\bm k}=(c_{\bm k\uparrow},c_{\bm k\downarrow})$ [$f_{\bm k}=(f_{\bm k\uparrow},f_{\bm k\downarrow})$] is the spinor for itinerant (localized) electrons,
\begin{align}
\varepsilon_{\bm k}=&-2t\left[\cos(k_x)+2\cos\left(\frac{k_x}2\right)\cos\left(\frac{\sqrt3k_y}2\right)\right]-\mu,\nonumber\\
\sim&-(6t+\mu)+\frac32tk^2,
\end{align}
is the eigenenergy of itinerant electrons on the triangular lattice with momentum $\bm k$, $k\equiv|\bm k|$, $\gamma_{i\bm k}\equiv e^{{\rm i}\bm k\cdot\bm r_i}$, $J>0$ is the Zeeman splitting of the localized electron, $\bm r_i$ is the position of $i$th spin, and $\vec S_i$ is a unit vector parallel to the magnetic moment of site $i$. Here, we assumed the site distance $a=1$. The eigenenergy of electrons are approximated by a quadratic dispersion. This model corresponds to a mean-field theory for the Anderson impurity model where the onsite interaction between the localized electrons are treated by Hartree-Fock approximation. Note that there is no spin-orbit interaction in Eq.~\eqref{eq:Hamil}.

We calculate the scattering rate $W_{\bm k\sigma,\bm k'\sigma'}$of electrons using $T$ matrix. The details of the derivation is elaborated in Materials and Methods section. We here summarize the main results we use in the rest of this paper. The $T$ matrix for the scattering by the spin cluster reads
\begin{align}
T_{\bm k\sigma,\bm k'\sigma'}=&\nonumber\\
\frac{V^2}{N}\sum_{i,j}&\gamma^\ast_{i\bm k}\gamma_{j\bm k'}\left[\frac1{\varepsilon+{\rm i}\delta+J\sum_l \bm S_l\cdot\bm\sigma_l-\Sigma(\varepsilon)}\right]_{i\sigma,j\sigma'},\label{eq:impurity:Tmatrix}
\end{align}
where $\Sigma(\varepsilon)$ is a matrix with its elements
\begin{align}
\Sigma_{i\sigma,j\sigma'}(\varepsilon)=&\frac{V^2}{4\pi^2}\delta_{\sigma\sigma'}\int d\bm k\, \frac{\gamma_{i\bm k}\gamma^\ast_{j\bm k}}{\varepsilon+{\rm i}\delta-\varepsilon_{\bm k}},
\end{align}
is the self-energy of localized electrons. The scattering rate is proportional to the square of $T$ matrix ${\cal W}_{\vec k\sigma,\vec k'\sigma'}\equiv |T_{\vec k\sigma,\vec k'\sigma'}|^2$,
\begin{align}
W_{\vec k\sigma,\vec k'\sigma'}=2\pi{\cal W}_{\vec k\sigma,\vec k'\sigma'}\delta(\varepsilon_{\vec k\sigma}-\varepsilon_{\vec k'\sigma'}).
\end{align}
This gives the scattering rate of electrons from the state with momentum $\bm k'$ and spin $\sigma'$ to that with $\bm k$ and $\sigma$.

We study the skew scattering by spin clusters using the average of ${\cal W}_{\bm k\sigma,\bm k'\sigma'}$ over the incident electron directions. We define the averaged ${\cal W}_{\bm k\sigma,\bm k'\sigma'}$ by
\begin{widetext}
\begin{align}
\bar{\cal W}_{\sigma,\sigma'}(\delta\phi)\equiv&\int\frac{d\phi'}{2\pi}{\cal W}_{\bm k\sigma,\bm k'\sigma'},\nonumber\\
=&\frac{V^4}{N^2}\sum_{i,j,m,n}\left[\frac1{J\sum_l \bm S_l\cdot\bm\sigma_l-\Sigma(\varepsilon)}\right]_{i\sigma,j\sigma'}\left[\frac1{J\sum_l \bm S_l\cdot\bm\sigma_l-\Sigma(\varepsilon)}\right]^\ast_{m\sigma,n\sigma'}\nonumber\\
&\hspace{2cm}\times J_0\left(k\sqrt{r_{jn}^2+r_{im}^2-2\bm r_{im}\cdot\bm r_{jn}\cos(\delta\phi)+2(\bm r_{im}\times\bm r_{jn})_z\sin(\delta\phi)}\right),\label{eq:result:Wfull}
\end{align}
\end{widetext}
where $\bm r_{ij}\equiv \bm r_i-\bm r_j$, $\phi'\equiv{\rm atan}(k'_y/k'_x)$ is the angle of incident electron, $\delta\phi$ is the difference of angles between the momentum of incoming and outgoing electrons, $\vec \sigma_l\equiv(\sigma_l^x,\sigma_l^y,\sigma_l^z)$ is a vector of matrix $\sigma_l^a\equiv E_{ll}\otimes\sigma^a$ ($a=x,y,z$ and $E_{ij}$ is the matrix unit), and $J_0(x)$ is the $n=0$ first Bessel function,
\begin{align}
J_0(x)=\sum_{n=0}^\infty \frac{(-1)^n}{(n!)^2}\left(\frac{x}{2}\right)^{2n}.
\end{align}
We define the averaged scattering rate calculated using Eq.~\eqref{eq:result:Wfull} by
\begin{align}
\bar W_{\sigma,\sigma'}(\delta\phi)=2\pi\bar{\cal W}_{\sigma,\sigma'}(\delta\phi)\delta(\varepsilon_{\vec k\sigma}-\varepsilon_{\vec k'\sigma'}).\label{eq:result:Wave}
\end{align}
Equations~\eqref{eq:result:Wfull} and \eqref{eq:result:Wave} gives the basis of our discussion in the rest of this work.

Equation~\eqref{eq:result:Wfull} implies the absence of skew scattering in one- and two-impurity cases. In the case of one impurity, $\bm r_{11}=0$. Therefore, $\bar{\cal W}_{\bm k\sigma,\bm k'\sigma'}$ has no $\delta\phi$ dependence. We can also show that the two-impurity cluster do not produce skew scattering. Suppose there are two impurities placed with a distance $r$; $r_{ij}=0$ if $i=j$ and $r_{ij}=r$ otherwise. According to Eq.~\eqref{eq:result:Wfull}, the angular dependence appears from the terms $i\ne m$ and $j\ne n$. In the two-impurity case, the product of two vectors are $\vec r_{im}\cdot\vec r_{jn}=\pm r^2$ and $(\vec r_{im}\times\vec r_{jn})_z=0$. By substituting $(\vec r_{im}\times\vec r_{jn})_z=0$ into Eq.~\eqref{eq:result:Wfull}, we obtain
\begin{widetext}
\begin{align}
\bar{\cal W}_{\vec k\sigma,\vec k'\sigma'}=&\frac{V^4}{N^2}\sum_{i,j,m,n}\left[\frac1{J\sum_l \vec S_l\cdot\vec\sigma_l-\Sigma(\varepsilon)}\right]_{i\sigma,j\sigma'}\left[\frac1{J\sum_l \vec S_l\cdot\vec\sigma_l-\Sigma(\varepsilon)}\right]^\ast_{m\sigma,n\sigma'}J_0\left(k\sqrt{r_{jn}^2+r_{im}^2-2\vec r_{im}\cdot\vec r_{jn}\cos(\delta\phi)}\right)
\end{align}
\end{widetext}
Therefore, $\bar{\cal W}_{\bm k\sigma,\bm k'\sigma'}=\bar{\cal W}'_{\sigma,\sigma'}(\delta\phi)$ is always symmetric with respect to $\delta\phi$. Namely, no skew scattering for the one- and two-impurity cases.

\vspace{.4cm}
\noindent
{\bf Giant skew scattering by a three-spin cluster\\}\hspace{.3cm}
The smallest spin cluster contributing to the skew scattering is the cluster with three spins. Previous studies finds the scattering by three-spin cluster causes skew scattering~\cite{Denisov2016,Ishizuka2017} and AHE~\cite{Tatara2002,Denisov2016,Ishizuka2017}. These theories are based on the perturbation expansion with respect to the Kondo coupling; the results are valid when the Kondo coupling is small compared to the Fermi energy. In contrast, we here study the behavior of electron scattering using a formalism which applies to arbitrary strength of electron-spin coupling.

\begin{figure}
  \includegraphics[width=\linewidth]{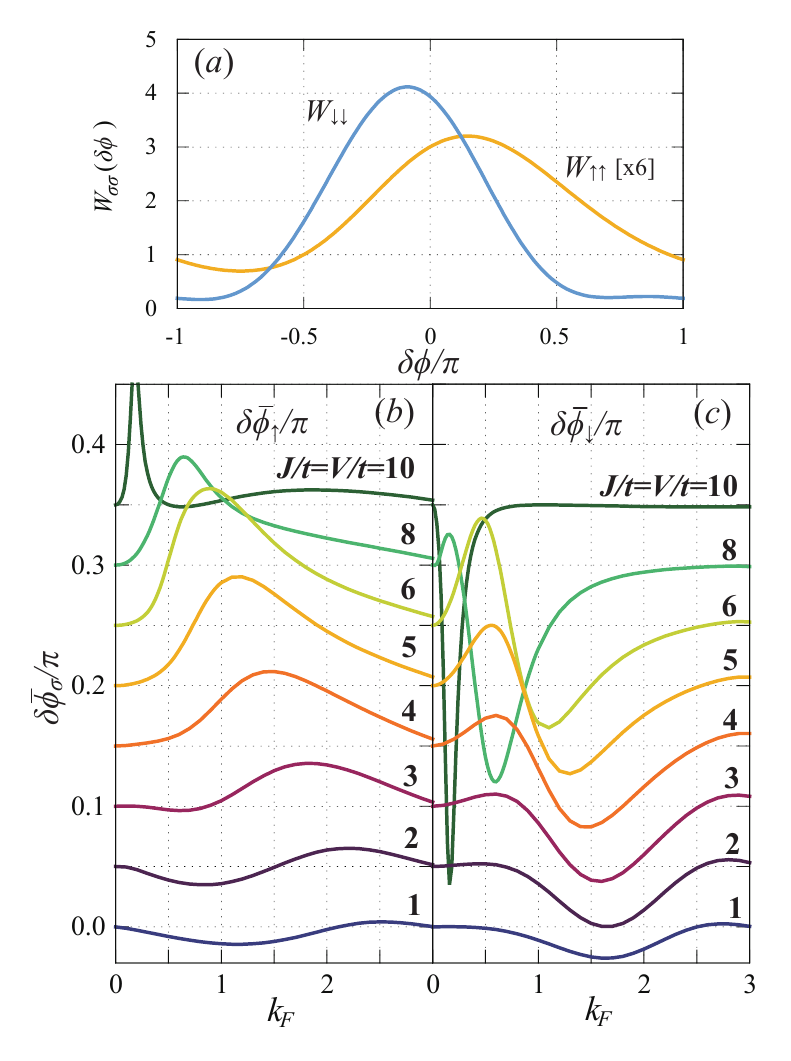}
  \caption{Numerical result of $\bar{\cal W}_{\bm k\sigma,\bm k'\sigma'}$ for a three-spin cluster. (a) $\delta\phi$ dependence of $\bar{\cal W}_{\uparrow,\uparrow}(\delta\phi)$ and $\bar{\cal W}_{\downarrow,\downarrow}(\delta\phi)$ for $J=V=6$, $k_F=1$, and $\theta=\pi/4$. (b,c) $k_F$ dependence of $\delta\bar\phi_\uparrow$ (b) and $\delta\bar\phi_\downarrow$ (c) for $\theta=\pi/4$. Different curves are for different $J$ and $V$.}
  \label{fig:result:Wkk}
\end{figure}

In this section, we consider a three-spin cluster consisting of three nearest-neighbor sites on the triangular lattice. We particularly focus on the umbrella configuration of spins where three spins are tilted by $\theta$ from the ferromagnetic configuration [Fig.~\ref{fig:intro:skew}(a)]. Figure~\ref{fig:result:Wkk}(a) shows the $\theta$ dependence of $\bar{\cal W}_{\uparrow,\uparrow}(\delta\phi)$ and $\bar{\cal W}_{\downarrow,\downarrow}(\delta\phi)$ for $J=V=6$ and $\theta=\pi/4$. The result is asymmetric with respect to $\delta\phi$ with the maximum of $\bar{\cal W}_{\uparrow,\uparrow}(\delta\phi)$ away from $\delta\phi=0$. This is a typical result of skew scattering, in which the scatterer scatters electrons asymmetrically.

The skewness of scattering is captured by the skew scattering angle
\begin{align}
\delta\bar\phi_{\sigma}=&\int_{-\pi}^\pi \frac{d(\delta\phi)}{\Omega_\sigma} \,\delta\phi\bar{\cal W}_{\sigma,\sigma}(\delta\phi),
\end{align}
where
\begin{align}
\Omega_\sigma=&\int_{-\pi}^\pi d(\delta\phi)\,\bar{\cal W}_{\sigma,\sigma}(\delta\phi).
\end{align} 
$\delta\bar\phi_\sigma$ is positive when the electrons are scattered rightward such as $\bar{\cal W}_{\uparrow,\uparrow}(\delta\phi)$ in Fig.~\ref{fig:result:Wkk}(a), and negative when scattered leftward as in $\bar{\cal W}_{\downarrow,\downarrow}(\delta\phi)$ in the same figure. Figures~\ref{fig:result:Wkk}(b) and \ref{fig:result:Wkk}(c) shows the Fermi wavenumber $k_F$ dependence of $\delta\bar\phi_\sigma$ for $J=V$ cases. We here set the cutoff $\Lambda=\pi$.

The results in Figs.~\ref{fig:result:Wkk}(b) and \ref{fig:result:Wkk}(c) shows distinct behaviors depending on $J/t,V/t$. The $J/t,V/t\ll1$ and $k_F\ll1$ case corresponds to the case studied in Ref.~\cite{Ishizuka2017}. $\delta\bar\phi_\uparrow$ and $\delta\bar\phi_\downarrow$ behaves similarly when $J/t=V/t=1$; the sign of $\delta\bar\phi_\sigma$ is negative for both spins with the minimum at around $k_F\sim1.5$. This is approximately consistent with the perturbation theory in Ref.~\cite{Ishizuka2017}, in which $\bar{\cal W}_{\uparrow,\uparrow}(\delta\phi)=\bar{\cal W}_{\downarrow,\downarrow}(\delta\phi)$. On the other hand, $\delta\bar\phi_\uparrow$ and $\delta\bar\phi_\downarrow$ generally behaves differently when $J/t, V/t$ is large. For instance, $\delta\bar\phi_\uparrow$ is always positive for $4\le J/t,V/t\le 8$ while $\delta\bar\phi_\downarrow$ shows oscillation in the sign. Overall, the sign of $\delta\bar\phi_\uparrow$ is positive and $\delta\bar\phi_\downarrow$ is negative when $J/t=V/t=10$. This is consistent with the double-exchange limit in which the coupling of localized moment and itinerant electrons produce fictitious magnetic field~\cite{Ye1999,Ohgushi2000}; the effective magnetic field for down spins has the opposite sign to the up spin. These results indicate the skew scattering shows a distinct behavior from the weak-coupling regime when $J/t,V/t$ is large.

Another important feature is the large skew angle. Figures~\ref{fig:result:Wkk}(b) and \ref{fig:result:Wkk}(c) shows a skew angle of order $\delta\bar\phi_\sigma={\cal O}(0.1\pi)$ when $J/t,V/t\gtrsim 4$. This is 10-100 times larger than the typical skew angle $\delta\bar\phi_\sigma\sim 10^{-3}\pi-10^{-2}\pi$ rad~\cite{Nagaosa2010}. This result implies that the skew scattering by the spin clusters produce a large AHE, which produces a large Hall angle in experiment.

\begin{figure}[tb]
  \includegraphics[width=\linewidth]{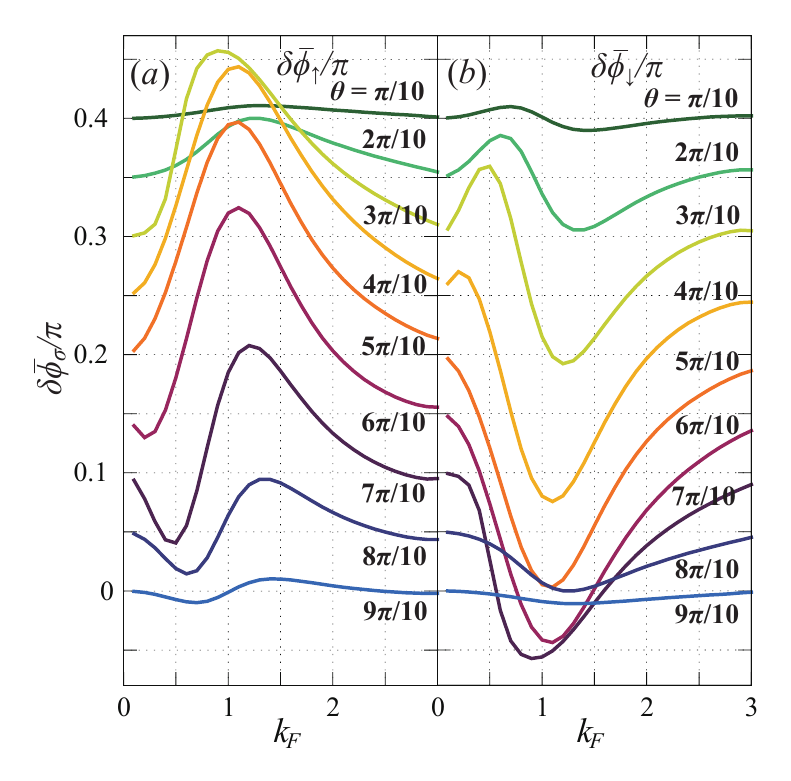}
  \caption{$k_F$ dependence of $\delta\bar\phi_\sigma$ for different $\theta$. The results for (a) $\delta\bar\phi_\uparrow$ and (b) $\delta\bar\phi_\downarrow$ for $J/t=V/t=5$. The transverse axis is $k_F$.}
  \label{fig:result:theta}
\end{figure}

The large skew angle generally appears in the three-spin cluster. We investigate this focusing on the $\theta$ dependence of $\delta\bar\phi_\sigma$. Figure~\ref{fig:result:theta} shows the $k_F$ dependence of the skew angle for different $\theta$ with $J/t=V/t=5$; Figure~\ref{fig:result:theta}(a) is for ${\cal W}_{\uparrow\uparrow}(\delta\phi)$ and Fig.~\ref{fig:result:theta}(b) is for ${\cal W}_{\downarrow\downarrow}(\delta\phi)$. The result shows $\delta\bar\phi_\sigma$ of order $0.1\pi$ rad when $\pi/5\le\theta\le4\pi/5$. Therefore, the thermally-fluctuating spins with local chiral correlation results in a large extrinsic anomalous Hall conductivity. 

\begin{figure}[tb]
  \includegraphics[width=\linewidth]{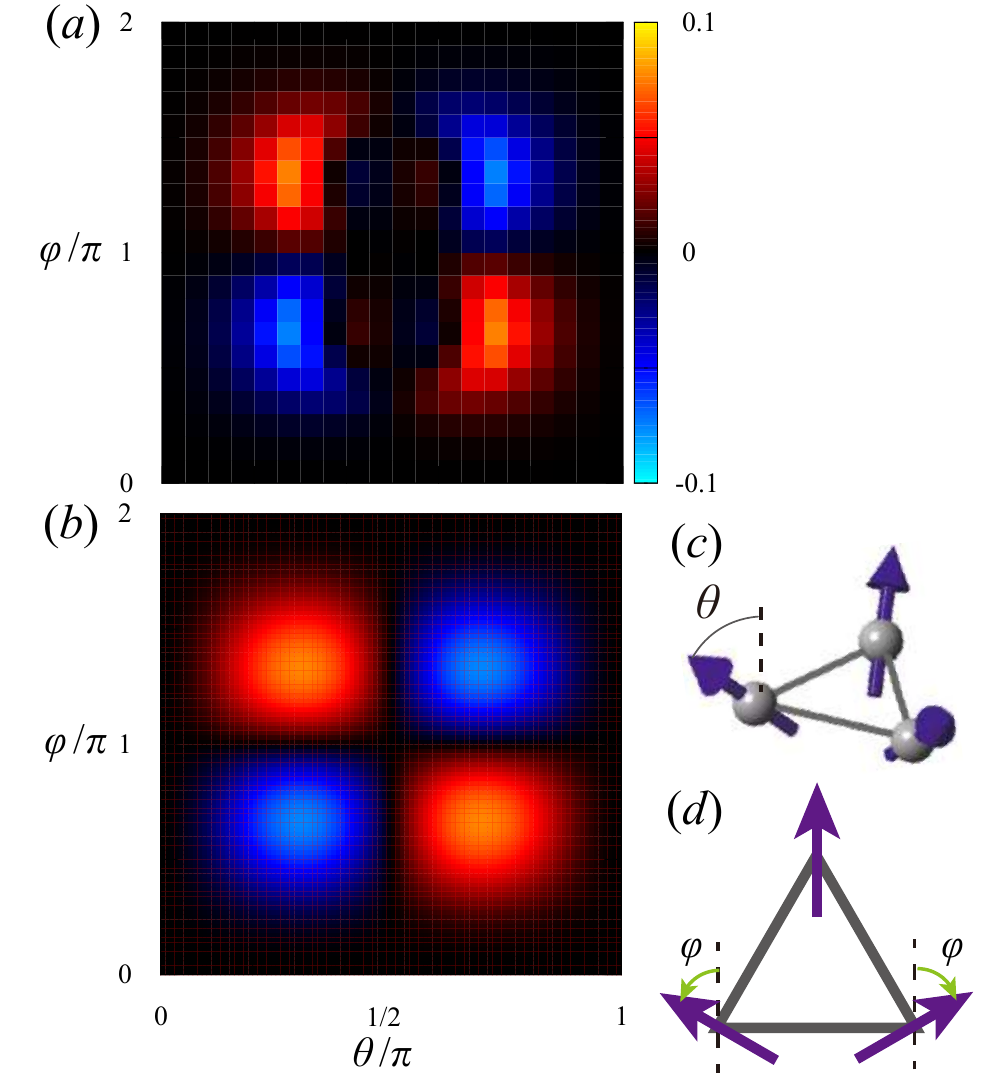}
  \caption{Spin-cluster scattering for coplanar spin textures. (a) Contour plot of $\delta\bar\phi^+/\pi$ calculated using Eq.~\eqref{eq:result:Wfull} and (b) the net scalar spin chirality with $\varphi_2=\pi/2$. (a) is the result for $J/t=V/t=6$, $k_F=1/2$, and $\theta=\pi/4$. $\theta$ is the canting angle as shown in (c). (d) shows the top view of (c). Here, $\varphi$ is the rotation of the in-plane component from the $y$ axis.}
  \label{fig:result:ahe}
\end{figure}

Despite the rich structure of ${\cal W}_{\sigma\sigma'}(\delta\phi)$ in Figs.~\ref{fig:result:Wkk} and \ref{fig:result:theta}, we find the average skew angle $\delta\bar\phi^{+}\equiv(\delta\bar\phi_{\uparrow}+\delta\bar\phi_{\downarrow})/2$ is approximately proportional to the scalar spin chirality. Figure~\ref{fig:result:ahe}(a) shows the contour plot of $\delta\bar\phi^{+}$ for $k_F=1/2$ and $J/t=V/t=6$; the plot is for canting angle $\theta$ and with the rotation $\varphi$ in the $xy$ plane [See Fig.~\ref{fig:result:ahe}(c) and \ref{fig:result:ahe}(d)]. The scalar spin chirality for the spin configuration is shown in Fig.~\ref{fig:result:ahe}(b) for comparison. Figures~\ref{fig:result:ahe}(a) and \ref{fig:result:ahe}(b) shares common features; they are both antisymmetric about $\theta=1/2$ and $\varphi=\pi$ lines, and the maximum in each quadrant is approximately at the same point. As $\delta\bar\phi^{+}$ is related to AHE, this result implies the close relation between AHE and the scalar spin chirality of spin cluster even when the coupling between electrons and spins is strong. We further discuss this aspect later based on the general property of ${\cal W}_{\sigma\sigma'}(\delta\phi)$.

The maximum of $|\delta\bar\phi_\sigma|$ is located at $k_F\sim 1$ in the current results. Namely, when the wavelength of the electrons is comparable to the distance between the spins. This feature resembles the scattering of magnons by skyrmions~\cite{Iwasaki2014}, where the maximum of skew angle is at a wavenumber comparable to the inverse of the diameter of the skyrmion. Reference~\cite{Iwasaki2014} also points out that their numerical simulation for the magnon scattering is well reproduced by the theory for Aharonov-Bohm scattering~\cite{Aharonov1959,Brown1985,Brown1987}. The current problem has a similar aspect to the magnon scattering when the coupling between the electrons and scatterers is strong; in this limit, the coupling of electrons and spins produces a fictitious magnetic field in a canted spin configuration~\cite{Ye1999,Ohgushi2000}. Hence, the enhancement at $k_F\sim1$ is most likely related to the inverse of the size of the spin cluster, which is $\sim 1$ in the current case.

The anomalous Hall effect due to cluster-spin scattering potentially results in an unconventional behavior in the scaling plot of the conductivities [Fig.~\ref{fig:intro:skew}(d)]. Within the relaxation-time approximation, it is known that the extrinsic anomalous Hall effect by skew scattering is proportional to the relaxation time while that by the intrinsic mechanism is insensitive. As a consequence, the skew scattering is  dominant in a clean material with high conductivity (larger relaxation time) while the intrinsic mechanism is dominant when the conductivity is low; the crossover typically occurs at a longitudinal conductivity $\sigma_{xx}\sim 10^5$ S/cm~\cite{Onoda2008}. This crossover applies to the typical case in which the skew scattering angle is $\sim10^{-3}\pi-10^{-2}\pi$ rad. On the other hand, the skew scattering by the spin clusters has the skew scattering angle of $\sim10^{-1}\pi$ rad, about 10-100 times larger than the conventional cases. As a consequence, the extrinsic Hall conductivity increases by 10-100 times for a given $\sigma_{xx}$. Therefore, the crossover shifts to a lower conductivity by 1-2 order of magnitude [Fig.~\ref{fig:intro:skew}(b)]. Therefore, the scaling plot of conductivities shows an unconventional plot if the spin-cluster scattering is dominant.

\vspace{.4cm}
\noindent
{\bf Extrinsic spin-Hall effect by the spin-cluster scattering\\}\hspace{.3cm}

\begin{figure}
\includegraphics[width=\linewidth]{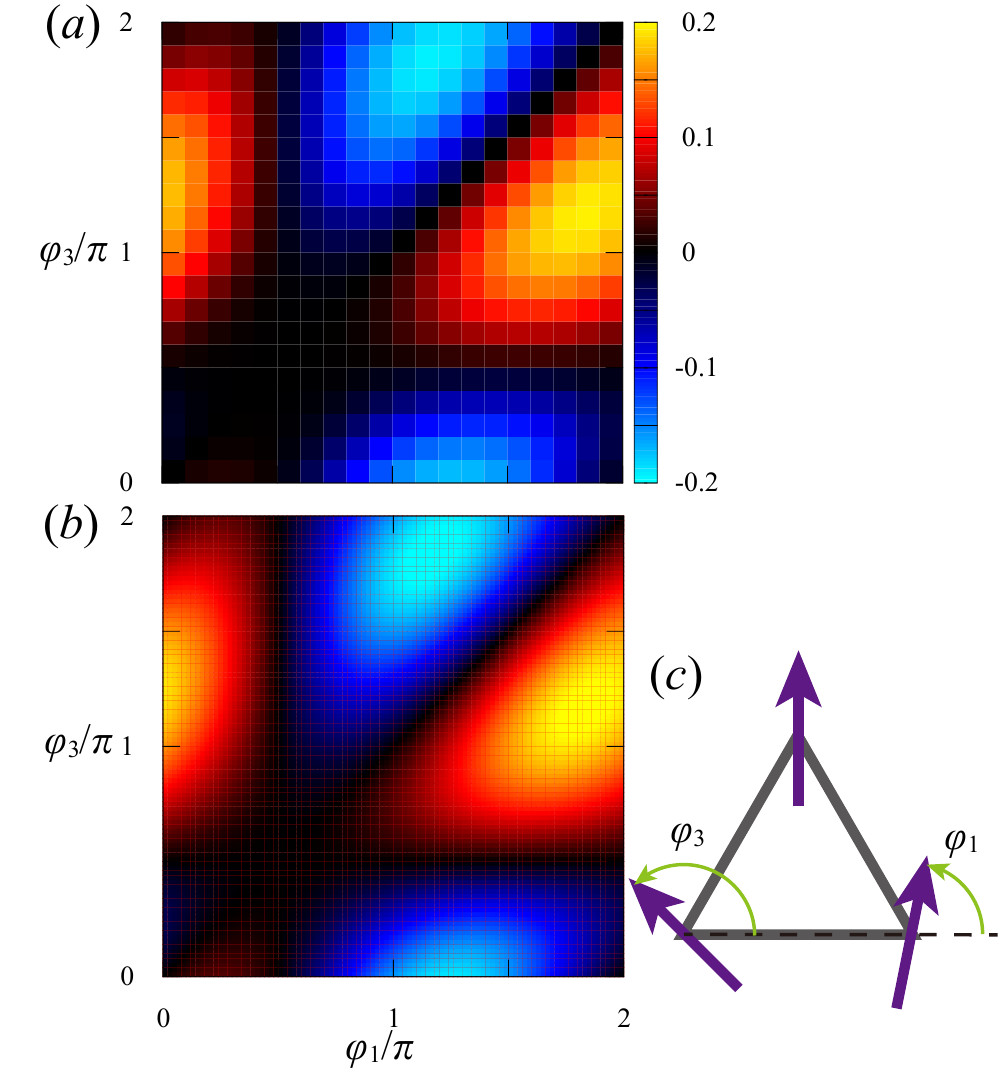}
\caption{Spin-cluster scattering for coplanar spin textures. (a) Contour plot of $\delta\bar\phi^-/\pi$ calculated using Eq.~\eqref{eq:result:Wfull} and (b) the net vector chirality $\chi_v$ in Eq.~\eqref{eq:vchirality} with $\varphi_2=\pi/2$. (a) is the result for $J/t=V/t=6$, $k_F=1$, and $\theta=\pi/2$. $\varphi_1$ and $\varphi_3$ are defined as in (c).}
\label{fig:she}
\end{figure} 

The skew scattering also causes spin Hall effect. In contrast to the anomalous Hall effect, the results for $\theta=5\pi/10(=\pi/2)$ in Fig.~\ref{fig:result:theta} implies a coplanar spin texture produces a finite spin Hall current; the skew angle for up and down spins has the opposite sign for arbitrary $k_F$.  Therefore, the transverse charge current cancels while that of the spin current remains finite. In this section, we study the spin dependent skew scattering focusing on the coplanar spin texture where all spins lie in the $xy$ plane.

Figure~\ref{fig:she}(a) shows the contour plot of $\delta\bar\phi^-$ when the three spins lies in the $xy$ plane; the result is for $V=J=t=1$. The two axis, $\varphi_1$ and $\varphi_3$ are the angle of two spins shown in Fig.~\ref{fig:she}(c). The result resembles that of the net vector chirality of three spins,
\begin{align}
\chi_v
=&\bm S_1\times\bm S_2+\bm S_2\times\bm S_3+\bm S_3\times\bm S_1,\nonumber\\
=&\sin(\varphi_2-\varphi_1)+\sin(\varphi_3-\varphi_2)+\sin(\varphi_1-\varphi_3),\label{eq:vchirality}
\end{align}
where $\varphi_2$ is the direction of $\bm S_2$ in Fig.~\ref{fig:she}(c). Figure~\ref{fig:she}(b) shows the contour plot of $\chi_v$ with $\varphi_2=\pi/2$.

On the other hand, single impurity spin and two-spin cluster do not produce a skew scattering in general. This fact is discussed in the above section. Therefore, a scattering process that involves three spins is necessary for a nonzero $\delta\bar\phi^-$. Intuitively, this is because we need at least three spins to define the plane in which the Hall effect takes place. Therefore, the three-spin cluster is necessary for a finite $\delta\bar\phi^-$.

To see the relation between the skew scattering and the spin chirality, we expand
\begin{align}
&\frac1{J\sum_l \vec S_l\cdot\vec\sigma_l-\Sigma(\varepsilon)}=\sum_{n=0}^\infty\left(-G_zH_f\right)^nG_z.
\end{align}
When $S_i^z=0$, the leading order term of $\bar{\cal W}_{\vec k,\vec k'}^-=(\bar{\cal W}_{\vec k\uparrow,\vec k'\uparrow}-\bar{\cal W}_{\vec k\downarrow,\vec k'\downarrow})/2$ appears from the third order in expansion; it reads
\begin{widetext}
\begin{align}
& \bar{\cal W}_{\vec k,\vec k'}^-\sim\frac{2V^4}{N^2}\sum_{i,m,n}{\rm Im}\left[\Sigma_{ii}\Sigma_{mn}^\ast\right](\bm S_m\times\bm S_n)_z J_0\left(k\sqrt{r_{im}^2+r_{in}^2-2\bm r_{im}\cdot\bm r_{in}\cos(\delta\phi)+2(\bm r_{im}\times\bm r_{in})_z\sin(\delta\phi)}\right).
\end{align}
Here, we used $\bm S_i\cdot\bm S_i=1$, which eliminates two spin variables from the above formula. In case of the three spin cluster, $\Sigma_{11}=\Sigma_{22}=\Sigma_{33}=\Sigma_d$ and $\Sigma_{12}=\Sigma_{23}=\Sigma_{31}=\Sigma_{od}$. Therefore, the above formula becomes
\begin{align}
& \bar{\cal W}_{\vec k,\vec k'}^-\sim\frac{2V^4}{N^2}{\rm Im}\left[\Sigma_{d}\Sigma_{od}^\ast\right]\sum_{m,n}(\bm S_m\times\bm S_n)_z \sum_iJ_0\left(k\sqrt{r_{im}^2+r_{in}^2-2\bm r_{im}\cdot\bm r_{in}\cos(\delta\phi)+2(\bm r_{im}\times\bm r_{in})_z\sin(\delta\phi)}\right).
\end{align}
\end{widetext}
The sum over $i$ is independent of $m$ and $n$. Therefore, the leading order in $V^2/(Jt)$ is proportional to the sum of the vector chirality $\bm S_i\times\bm S_j$ while it requires (at least) three spins.

\begin{figure}
  \includegraphics[width=0.9\linewidth]{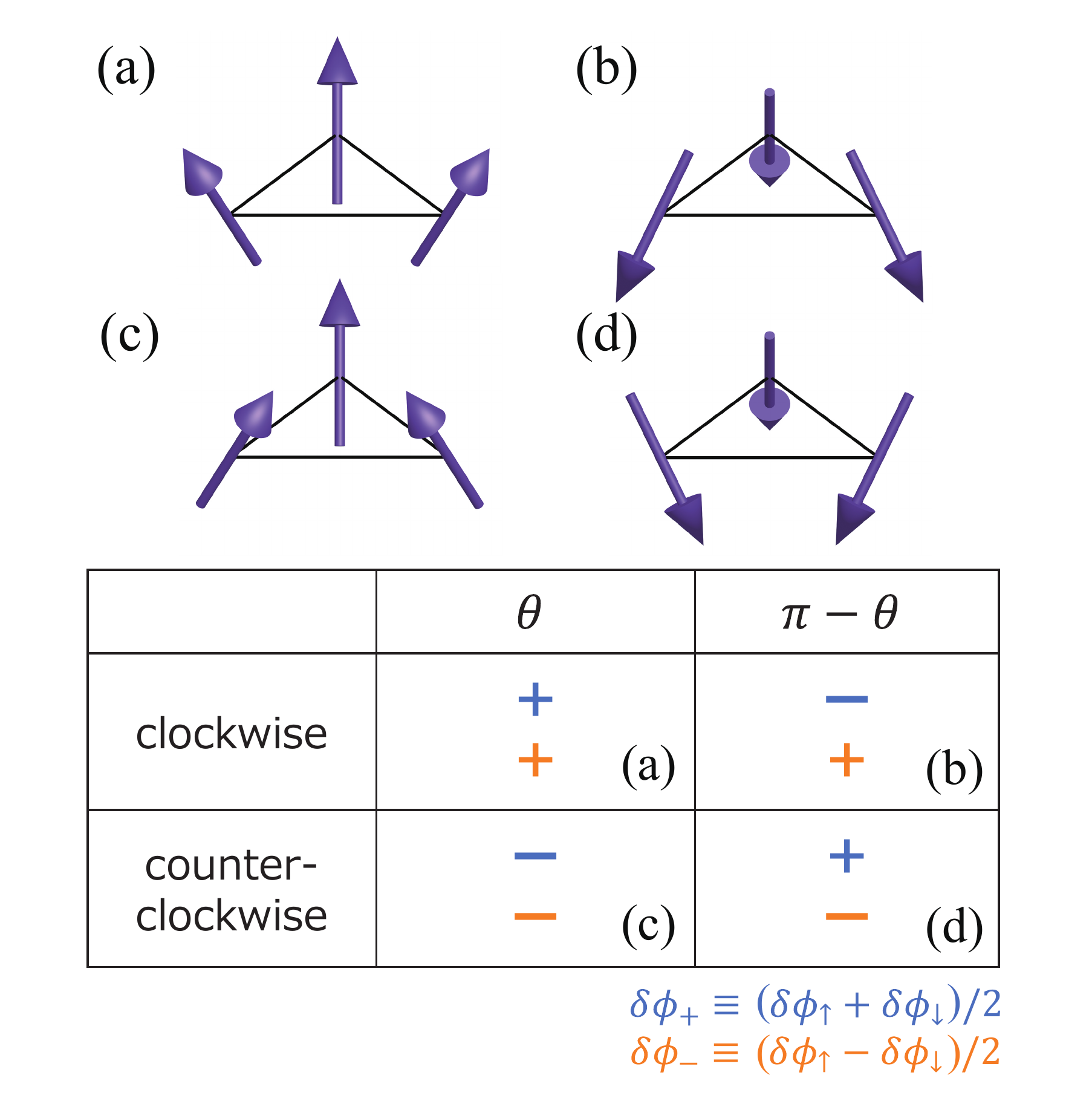}
  \caption{The relation of the sign of $\delta\bar\phi_+$ (related to anomalous Hall effect) and  $\delta\bar\phi_-$ (spin Hall effect). (a-d) Four spin configurations which show different signs of $\delta\bar\phi_\pm$: The clockwise (a,b) or counter-clockwise (c,d) orientation and the canting angle $\theta$ (a,c) or $\pi-\theta$ (b,d). The sign of $\delta\bar\phi_\pm$ are summarized in the bottom table. The upper sign in each block is for $\delta\bar\phi_+$ and the lower one is for $\delta\bar\phi_-$. The alphabet in each cell shows corresponding spin configuration in (a-d).}
  \label{tab:result:sigmaxy}
\end{figure}

\vspace{.4cm}
\noindent
{\bf Spin-cluster scattering and spin chirality\\}\hspace{.3cm}
The above results show that the spin-cluster scattering produces rich behaviors in the scattering phenomena. We here organize the relation between the spin configurations studied above and the anomalous/spin Hall coefficients. The discussion here is based on the three general properties of scattering rate $\bar W_{\sigma,\sigma'}(\delta\phi)$. The results are summarized in Fig.~\ref{tab:result:sigmaxy}. We find that the sign of skew angle changes depending on the orientation of spins (clockwise or anti-clockwise) and the canting angle $\theta$ or $\pi-\theta$; four spin configurations with different orientation and canting angle are shown in Figs.~\ref{tab:result:sigmaxy}(a)-\ref{tab:result:sigmaxy}(d).

The table in Fig.~\ref{tab:result:sigmaxy} is obtained from the properties of $\bar W_{\sigma,\sigma'}(\delta\phi)$, which are explained below. In the table, we considered $\delta\bar\phi_\pm=(\delta\bar\phi_\uparrow\pm\delta\bar\phi_\downarrow)/2$ instead of $\delta\bar\phi_\sigma$ because they are directly related to extrinsic anomalous ($\delta\bar\phi_+$) and spin ($\delta\bar\phi_-$) Hall effects.

\noindent{\it 1. $\bar W_{\sigma,\sigma'}(\delta\phi)|_{\theta}=\bar W_{\sigma,\sigma'}(\delta\phi)|_{-\theta}$.} --- Here, $\bar W_{\sigma,\sigma'}(\delta\phi)|_{\theta}$ is the scattering rate for the three spin cluster with canting angle $\theta$; the spins cant outward [Fig.~\ref{fig:intro:skew}(b)] when $\theta>0$ and inward [Fig.~\ref{fig:intro:skew}(c)] when $\theta<0$. The relation is explicitly shown by rewriting Eq.~\eqref{eq:result:Wfull}. We expand the Green function in Eq.~\eqref{eq:result:Wfull},
\begin{align}
&\frac1{J\sum_l \vec S_l\cdot\vec\sigma_l-\Sigma(\varepsilon)}=\nonumber\\
&\qquad\sum_{n=0}^\infty\left(G_zH'\right)^{2n}G_z-\sum_{n=0}^\infty\left(G_zH'\right)^{2n}G_zH'G_z,
\end{align}
where $G_z=[J\sum_l S_l^z\sigma_l^z-\Sigma(0)]^{-1}$ and $H'=\sum_lS_l^x\sigma_l^x+S_l^y\sigma_l^y$. The first term of this equation is diagonal in the spin index while the diagonal elements in the second terms are zero. Substituting this formula into Eq.~\eqref{eq:result:Wfull}, we find
\begin{widetext}
\begin{subequations}
\begin{align}
\bar{\cal W}_{\vec k\sigma,\vec k'\sigma'}=&\frac{V^4}{N^2}\sum_{i,j,m,n}\left[\left(G_zH'\right)^{2n}G_z\right]_{i\sigma,j\sigma'}\left[\left(G_zH'\right)^{2n}G_z\right]^\ast_{m\sigma,n\sigma'}\nonumber\\
&\hspace{2cm}\times J_0\left(k\sqrt{r_{jn}^2+r_{im}^2-2\vec r_{im}\cdot\vec r_{jn}\cos(\delta\phi)+2(\vec r_{im}\times\vec r_{jn})_z\sin(\delta\phi)}\right),
\end{align}
for $\sigma=\sigma'$ and
\begin{align}
\bar{\cal W}_{\vec k\sigma,\vec k'\sigma'}=&\frac{V^4}{N^2}\sum_{i,j,m,n}\left[\sum_{n=0}^\infty\left(G_zH'\right)^{2n}G_zH'G_z\right]_{i\sigma,j\sigma'}\left[\sum_{n=0}^\infty\left(G_zH'\right)^{2n}G_zH'G_z\right]^\ast_{m\sigma,n\sigma'}\nonumber\\
&\hspace{2cm}\times J_0\left(k\sqrt{r_{jn}^2+r_{im}^2-2\vec r_{im}\cdot\vec r_{jn}\cos(\delta\phi)+2(\vec r_{im}\times\vec r_{jn})_z\sin(\delta\phi)}\right),
\end{align}
\end{subequations}
\end{widetext}
for $\sigma\ne\sigma'$. Therefore $\bar W_{\sigma,\sigma'}(\delta\phi)\to \bar W_{\sigma,\sigma'}(\delta\phi)$ /because $\theta\to-\theta$ transforms $G_z\to G_z$ and $H'\to -H'$; the scattering rate does not change. This is consistent with the conventional notion because the transformation $\theta\to-\theta$ neither changes scalar or vector spin chiralities.

\vspace{4mm}
\noindent{\it2. $\bar W_{\sigma,\sigma'}(\delta\phi)|_{c}= \bar W_{\sigma,\sigma'}(-\delta\phi)|_{cc}$.} --- Here, $\bar W_{\sigma,\sigma'}(\delta\phi)|_{c}$ and $\bar W_{\sigma,\sigma'}(-\delta\phi)|_{cc}$ are respectively the scattering rate for clockwise and counter-clockwise configurations. The relation implies the Hall conductivity switches the sign by changing the sign of chirality. Formally, the clockwise to counter-clockwise transformation is equivalent to switching the positions of two sites, e.g., $\bm r_1\leftrightarrow\bm r_3$. We define the switched positions by $\bm r'_l$:
\begin{align}
\bm r_1 =\bm r'_3,\qquad\bm r_2 =\bm r'_2,\qquad\bm r_3 =\bm r'_1.
\end{align}
For the particular choice of $\bm r_l$,
\begin{align}
\bm r_1=(-1/2,0),\quad\bm r_2=(0,\sqrt3/2),\quad\bm r_3=(1/2,0),
\end{align}
the transposition $\bm r_i\to\bm r'_i$ is equivalent to the mirror operation about $x$ axis: $x\to -x$ and $y\to y$. Therefore,
\begin{align}
\bm r_{ij}\cdot\bm r_{nm} = \bm r'_{ij}\cdot\bm r'_{nm},\quad \bm r_{ij}\times\bm r_{nm} =-\bm r'_{ij}\times\bm r'_{nm}.
\end{align}
Therefore, the scattering rate after the transformation reads
\begin{widetext}
\begin{align}
\bar{\cal W}'_{\sigma,\sigma'}(\delta\phi)=&\frac{V^4}{N^2}\sum_{i,j,m,n}\left[\frac1{J\sum_l \bm S_l\cdot\bm\sigma_l-\Sigma(\varepsilon)}\right]_{i\sigma,j\sigma'}\left[\frac1{J\sum_l \bm S_l\cdot\bm\sigma_l-\Sigma(\varepsilon)}\right]^\ast_{m\sigma,n\sigma'}\nonumber\\
&\hspace{1.5cm}\times J_0\left(k\sqrt{r_{jn}^2+r_{im}^2-2\bm r_{im}\cdot\bm r_{jn}\cos(-\delta\phi)+2(\bm r_{im}\times\bm r_{jn})_z\sin(-\delta\phi)}\right),\\
=&\bar{\cal W}_{\sigma,\sigma'}(-\delta\phi).\nonumber
\end{align}
\end{widetext}
This transformation changes both the scalar and the vector spin chiralities. In the view of $\delta\bar\phi_\pm$, the above result shows both $\delta\bar\phi_+$ and  $\delta\bar\phi_-$ for counter-clockwise configuration has the opposite sign to that of the clockwise configuration [see the table in Fig.~\ref{tab:result:sigmaxy}].

\vspace{4mm}
\noindent{\it3. $\bar W_{\sigma,\sigma'}(\delta\phi)|_\theta = \bar W_{\bar\sigma,\bar\sigma'}(-\delta\phi)|_{\pi-\theta}$.} --- Here, $\bar\sigma=\downarrow,\uparrow$ for $\sigma=\uparrow,\downarrow$. The relation is implied from the $\pi$ rotation about an axis parallel to the incident momentum $\bm k'$ [Fig.~\ref{fig:intro:skew}(b)]. Suppose the incident momentum is parallel to the solid line in Fig.~\ref{fig:intro:skew}(b). Then, the $\pi$ rotation about this axis and $\varphi_R$ rotation about the axis perpendicular to the triangle transforms the spin cluster with $\theta$ to the cluster with $\pi-\theta$. This transformation indicates that there is a relation between $W_{\bm k\sigma,\bm k'\sigma'}|_\theta$ and $W_{\bm {\tilde k}\bar\sigma,\bm {\tilde k}'\bar\sigma'}|_{\pi-\theta}$. Here, the wavenumbers of incoming and outgoing electrons for $\pi-\theta$ configuration is not necessarily the same as $\bm k'$ and $\bm k$, which is represented by tilde. However, the $\pi$ rotation gives relations $\bm k'\cdot\bm k=\bm {\tilde k}'\cdot\bm {\tilde k}$ and $\bm k'\times\bm k=-\bm {\tilde k}'\times\bm {\tilde k}$; there is a relation between the rate of electrons scattered to one side in the $\theta$ configuration and the rate to the opposite side in $\pi-\theta$ configuration. This relation implies $\bar W_{\sigma,\sigma'}(\delta\phi)|_\theta = \bar W_{\bar\sigma,\bar\sigma'}(-\delta\phi)|_{\pi-\theta}$ because we take sum over all directions for the incident $\bm k$. This transformation changes the scalar spin chirality but not the vector spin chirality. Regarding $\delta\bar\phi_\pm$, the above transformation changes the sign of $\delta\bar\phi_+$ while it leaves  $\delta\bar\phi_-$ invariant [see the table in Fig.~\ref{tab:result:sigmaxy}]. 
 
The results obtained from the above arguments are summarized in Fig.~\ref{tab:result:sigmaxy}. In this table, each of the four blocks corresponds to different pair of signs for the scalar chirality and the $z$ component of vector spin chirality; the counter-clockwise configurations have the opposite sign of both scalar and vector chiralities compared to the clockwise ones, and $\pi-\theta$ configurations have the same vector spin chirality and opposite scalar spin chirality. Suppose we define the sign of both scalar and vector chiralities positive for the clockwise $\theta$ configuration. Then the scalar spin chirality is positive for clockwise $\theta$ and counter-clockwise $\pi-\theta$ configurations. On the other hand, the $z$ component of the vector chirality is positive for  the two clockwise configurations. As shown in the table of Fig.~\ref{tab:result:sigmaxy}, the sign of $\delta\bar\phi_+$ obeys that of the scalar spin chirality while $\delta\bar\phi_-$ follows that of the vector spin chirality. The result indicates the close relation between the spin chiralities and AHE/SHE despite the rich features seen in Figs.~\ref{fig:result:Wkk} and \ref{fig:result:theta}, e.g., sign change of $\delta\bar\phi_\sigma$ by changing $k_F$, $J/t$, and $V/t$.

This argument is consistent with the results in Figs.~\ref{fig:result:ahe} and \ref{fig:she}. In the two figures, we find that the contour plot of $\delta\bar\phi^+$ ($\delta\bar\phi^-$) resembles that of the scalar (vector) spin chirality. The above argument shows that the symmetry of $\delta\bar\phi^\pm$ corresponds to that of corresponding spin chiralities. Therefore, the spin configuration dependence of $\delta\bar\phi^\pm$ should look similar to that of the corresponding chiralities.

\section{Discussions}

To summarize, in this work, we systematically studied the skew scattering of electrons by three-spin clusters. Using an Anderson impurity model and the Green-function method, we calculated the scattering rate of the spin clusters for an arbitrary strength of the impurity-spin electron coupling. We find spin cluster causes a skew scattering with a large skew scattering angle in the order of $0.1\pi$ rad; this is 10-100 times larger than the typical skew scattering by non-magnetic impurities. This cluster skew scattering potentially produces a large anomalous and spin Hall effects related to the local spin correlation. When the cluster skew scattering is dominant, the scaling relation~\cite{Onoda2008} of the longitudinal and transverse conductivities deviates from the scaling plot, as shown in Fig.~\ref{fig:intro:skew}(d). These results show that the cluster skew scattering with strong coupling shows rich behaviors different from that in the weak-coupling limit. 

Regarding the experiments, a recent experiment on the Hall effect of MgZnO/ZnO thin films finds a large Hall angle of order $0.1\pi$ rad~\cite{Maryenko2017}; the anomalous Hall conductivity scales linearly with the longitudinal conductivity. The origin of the Hall effect is not clear. However, it was discussed that the magnetic moments in ZnO plays a role. As the physics takes place in the interface between MgZnO and ZnO, the symmetry breaking by the interface possibly produces the interfacial Dzyaloshinskii-Moriya interaction. The Dzyaloshinskii-Moriya interaction then induces chiral spin correlation under the external magnetic field. Hence, the cluster skew scattering discussed here should take place in this material.

In a different experiment, a large spin Hall effect was recently reported in Pd- and Au-based metallic spin glasses above the spin-glass transition temperature~\cite{Jiao2018}. In these materials, the effective exchange interactions between the spins are believed to be mediated by itinerant electrons, i.e., Ruderman-Kittel-Kasuya-Yosida (RKKY) interaction~\cite{Ruderman1954,Kasuya1956,Yosida1957}. As the typical length scale of RKKY interaction is given by $1/k_F$, the spin correlation typically has a structure of $1/k_F$. On the other hand, our results above show that the skew scattering is enhanced when the magnetic structure has a size of $\sim1/k_F$. Therefore, the RKKY interaction tunes the magnetic configuration to that produce a large skew scattering and the extrinsic spin Hall effect. This result implies the metallic spin glass is an ideal material for realizing the large extrinsic spin Hall effect.

\section{Materials and Methods}

\noindent
{\bf $T$ matrix of the magnetic-impurity model\\}\hspace{.3cm}
We here review a Green's function formula for calculating $T$ matrix, which is convenient for our study. A similar technique was used to study Anderson impurity models~\cite{Hewson1993}. The formula applies to a general system with two subspaces $A$ and $B$; the size of the Hilbert spaces are $N_A$ and $N_B$ for $A$ and $B$, respectively. For the sake of convenience, we note the $N_A\times N_A$ matrix Green function for $A$ subspace as $G_A$ and that for $B$ as $G_B$; the $N_A\times N_B$ matrix corresponding to the inter-subspace Green function elements of $A$ and $B$ is $G_{AB}$ and the other inter-subspace elements is $G_{BA}$.

We calculate the $T$ matrix from the Green function. The Dyson equation for Green function reads
\begin{align}
(\varepsilon\pm{\rm i}\delta-H_A)G_A^\pm-H'_{AB}G_{BA}^\pm=&1,\label{eq:Tmatrix:DysonAA}\\
(\varepsilon\pm{\rm i}\delta-H_B)G_B^\pm-H'_{BA}G_{AB}^\pm=&1,\\
(\varepsilon\pm{\rm i}\delta-H_A)G_{AB}^\pm-H'_{AB}G_{B}^\pm=&0,\\
(\varepsilon\pm{\rm i}\delta-H_B)G_{BA}^\pm-H'_{BA}G_{A}^\pm=&0.
\end{align}
Here, $H_A$ and $H_B$ are the Hamiltonian matrix within each subspace and $H'_{AB}$ and $H'_{BA}$ are the Hamiltonian elements that connects $A$ and $B$ subspaces. The last equation implies
\begin{align}
G_{BA}^\pm=&G_B^{0\pm}H'_{BA}G_{A}^\pm,
\end{align}
where
\begin{align}
G^{0\pm}_B=\frac1{\varepsilon\pm{\rm i}\delta-H_B},
\end{align}
is the Green function for the decoupled $B$ subspace (when $H'_{AB}=H'_{BA}=0$). Substituting this result to Eq.~\eqref{eq:Tmatrix:DysonAA}, $G_A$ reads
\begin{align}
G_A^\pm=&\frac1{(G_A^{0\pm})^{-1}-H'_{AB}G_B^{0\pm}H'_{BA}},
\end{align}
and hence
\begin{align}
G_{BA}^\pm=&G_B^{0\pm}H'_{BA}\frac1{(G_A^{0\pm})^{-1}-H'_{AB}G_B^{0\pm}H'_{BA}}.
\end{align}
Similarly, we find
\begin{align}
G_B^\pm=&\frac1{(G_B^{0\pm})^{-1}-H'_{BA}G_A^{0\pm}H'_{AB}},
\end{align}
and hence
\begin{align}
G_{AB}^\pm=&G_A^{0\pm}H'_{AB}\frac1{(G_B^{0\pm})^{-1}-H'_{BA}G_A^{0\pm}H'_{AB}}.
\end{align}
Using the general property of adjoint matrices, $(A^\dagger)^{-1}=(A^{-1})^\dagger$, $G^\pm_{AB}$ reads
\begin{align}
G_{AB}^\pm=&(G_{BA}^\mp)^\dagger,\\
=&\frac1{(G_A^{0\pm})^{-1}-H'_{AB}G_B^{0\pm}H'_{BA}} H'_{AB}G_B^{0\pm},
\end{align}
and
\begin{align}
G_B^\pm
=&G_B^{0\pm}+G_B^{0\pm}H'_{BA}\frac1{(G_A^{0\pm})^{-1}-H'_{AB}G_B^{0\pm}H'_{BA}} H'_{AB}G_B^{0\pm}.\label{eq:Tmatrix:GB}
\end{align}
Here, we defined the decoupled Green function for $A$ ($G_A^0$) in a similar manner to $G_B^0$. The comparison of Eq.~\eqref{eq:Tmatrix:GB} to the $T$ matrix representation, $G_B=G_B^0+G_B^0 T G_B^0$, implies
\begin{align}
T =&H'_{BA}\frac1{(G_A^{0\pm})^{-1}-H'_{AB}G_B^{0\pm}H'_{BA}} H'_{AB}.\label{eq:Tmatrix:T}
\end{align}
This is the general formula for the $T$ matrix of $B$ subspace treating $A$ as the scatterer.

\vspace{.4cm}
\noindent
{\bf Averaged scattering rate\\}\hspace{.3cm} 
The skew scattering by spin cluster is studied focusing on the scattering rate
\begin{align}
W_{\bm k\sigma,\bm k'\sigma'}\equiv 2\pi{\cal W}_{\bm k\sigma,\bm k'\sigma'}\delta(\varepsilon_{\bm k\sigma}-\varepsilon_{\bm k'\sigma'}),\label{eq:averagedW:Wdef}
\end{align}
where ${\cal W}_{\bm k\sigma,\bm k'\sigma'}\equiv |T_{\bm k\sigma,\bm k'\sigma'}|^2$, $\bm k'$ and $\bm k$ are the wavenumbers of incomming and outgoing waves, and $\varepsilon_{\bm k\sigma}$ is the eigenenergy of the electrons with momentum $\bm k$ and spin $\sigma=\uparrow,\downarrow$. In the main text, we focused on the paramagnetic case in which $\varepsilon_{\bm k\sigma}=\varepsilon_{\bm k}$. This quantity shows the rate of electron scattering from the states with momentum $\bm k'$ and $\sigma'$ to that with $\bm k$ and $\sigma$. The delta function in Eq.~\eqref{eq:averagedW:Wdef} reflects the scattering is an elastic one; this is because we treat the magnetic moment within the mean-field approximation. The skew scattering of electrons is manifested in the asymmetry of ${W}_{\bm k\sigma,\bm k'\sigma'}$, that is, $W_{\bm k\sigma,\bm k'\sigma'}\ne W_{\bm k'\sigma',\bm k\sigma}$.

The skew scattering is studied by considering the averaged ${\cal W}_{\bm k\sigma,\bm k'\sigma'}$ over the incident wave direction. ${\cal W}_{\bm k\sigma,\bm k'\sigma'}$ for the magnetic impurity clusters reads,
\begin{align}
&{\cal W}_{\bm k\sigma,\bm k'\sigma'}
=\frac{V^4}{N^2}\sum_{i,j,m,n}e^{{\rm i}\bm k'\cdot(\bm r_j-\bm r_n)-{\rm i}\bm k\cdot(\bm r_i-\bm r_m)}\times\nonumber\\
&\left[\frac1{J\sum_l \bm S_l\cdot\bm\sigma_l-\Sigma(\varepsilon)}\right]_{i\sigma,j\sigma'}\left[\frac1{J\sum_l \bm S_l\cdot\bm\sigma_l-\Sigma(\varepsilon)}\right]^\ast_{m\sigma,n\sigma'}.
\end{align}
The average of ${\cal W}_{\bm k\sigma,\bm k'\sigma'}$ is calculated by a substitution $\bm k=k(\cos(\phi'+\delta\phi),\sin(\phi'+\delta\phi))$ and $\bm k'=k(\cos(\phi'),\sin(\phi'))$ and calculating the average over $\phi'$. With this procedure, we find
\begin{widetext}
\begin{align}
\bar{\cal W}_{\bm k\sigma,\bm k'\sigma'}\equiv&\int\frac{d\phi'}{2\pi}{\cal W}_{\bm k\sigma,\bm k'\sigma'},\\
=&\frac{V^4}{N^2}\sum_{i,j,m,n}\left[\frac1{J\sum_l \bm S_l\cdot\bm\sigma_l-\Sigma(\varepsilon)}\right]_{i\sigma,j\sigma'}\left[\frac1{J\sum_l \bm S_l\cdot\bm\sigma_l-\Sigma(\varepsilon)}\right]^\ast_{m\sigma,n\sigma'}\nonumber\\
&\hspace{2cm}\times J_0\left(k\sqrt{r_{jn}^2+r_{im}^2-2\bm r_{im}\cdot\bm r_{jn}\cos(\delta\phi)+2(\bm r_{im}\times\bm r_{jn})_z\sin(\delta\phi)}\right),
\end{align}
\end{widetext}
where $\bm r_{ij}\equiv \bm r_i-\bm r_j$, and $J_0(x)$ is the $n=0$ first Bessel function,
\begin{align}
J_0(x)=\sum_{n=0}^\infty \frac{(-1)^n}{(n!)^2}\left(\frac{x}{2}\right)^{2n}.
\end{align}
This formula is used to discuss the skew scattering by the impurity clusters.

\vspace{.4cm}
\noindent
{\bf Acknowledgements:}\hspace{.2cm} We thank Y. Fujishiro, N. Kanazawa, D. Maryenko, and Y. Tokura for fruitful discussions. This work was supported by JSPS KAKENHI Grant Numbers JP18H04222, JP18H03676, and JP19K14649, and JST CREST Grant Numbers JPMJCR16F1 and JPMJCR1874, Japan. 
{\bf Author contributions:}\hspace{.2cm} H.I. and N.N. contributed equally to the $T$ matrix calculation and the analysis of the results. The manuscript was prepared by the two authors. N.N. supervised the project. {\bf Competing interests:}\hspace{.2cm} The authors declare no competing interests. {\bf Data and materials availability:}\hspace{.2cm} All data used to obtain the conclusions in this paper are presented in the paper. Other data may be requested from the authors.
\end{document}